\documentclass[onecolumn,aps,superscriptaddress]{revtex4-2}
\usepackage[dvips]{graphicx}
\usepackage{latexsym,amssymb,amsmath}
\usepackage{color}
\usepackage{bm}
\usepackage{enumerate}
\usepackage[bookmarksnumbered,bookmarksopen,colorlinks,citecolor=blue,linkcolor=blue]{hyperref}
\usepackage{mathrsfs}
\usepackage{times}
\usepackage{csquotes}
\usepackage{multirow}
\usepackage{feynmp}
\usepackage{tikz}

\graphicspath{{./figs/}}

\begin{document}   

\title{Suppressed Drell-Yan process by an external magnetic field}
\author{Shile Chen}
\email{csl2023@mail.tsinghua.edu.cn}
\affiliation{Department of Physics, Tsinghua University, Beijing 100084, China}
\author{Jiaxing Zhao}
\email{jzhao@itp.uni-frankfurt.de}
\affiliation{Helmholtz Research Academy Hesse for FAIR (HFHF), GSI Helmholtz Center for Heavy Ion Physics, Campus Frankfurt, 60438 Frankfurt, Germany}
 \affiliation{Institut f\"ur Theoretische Physik, Johann Wolfgang Goethe-Universität,Max-von-Laue-Straße 1, D-60438 Frankfurt am Main, Germany}
\author{Pengfei Zhuang}
\email{zhuangpf@mail.tsinghua.edu.cn}
\affiliation{Department of Physics, Tsinghua University, Beijing 100084, China}
\date{\today}

\begin{abstract}
The strongest electromagnetic fields in nature are created in high energy nuclear collisions and expected to change the dynamic scattering processes in the early stage. The magnetic field effect on the Drell-Yan process is investigated in this work. The single photon decay into quark pairs and lepton pairs in an external magnetic field leads to a significant Drell-Yan suppression in low and intermediate invariant mass region. The calculation up to the Landau level $n-1$ is complete in the energy region $s<2/3n(eB)$, and the enlarged phase space at higher landau levels may enhance the dilepton spectrum in the high mass region.   
\end{abstract}
\maketitle 

\section{Introduction}
Very Strong electromagnetic fields are predicted to be created in non-central relativistic heavy-ion collisions~\cite{Kharzeev:2007jp,Skokov:2009qp,Voronyuk:2011jd,Deng:2012pc,Tuchin:2013ie}. The maximum of the magnetic field can reach $5\ m_{\pi}^2\sim 0.1\ \rm GeV^2$ in Au+Au collisions at top RHIC energy and almost $70\ m_\pi^2\sim 1\ \rm GeV^2$ in Pb+Pb collisions at LHC energies~\cite{Deng:2012pc,Tuchin:2013ie}, where $m_\pi$ is the pion mass. 
The strong electromagnetic fields provide us a chance to search for novel transport phenomena such as the chiral magnetic effect~\cite{Kharzeev:2007jp,Fukushima:2008xe}, the splitting of $D^0$ and $\bar D^0$ directed flow~\cite{Das:2016cwd,STAR:2019clv,ALICE:2019sgg} and spin-polarized difference between $\Lambda$ and $\bar \Lambda$~\cite{Muller:2018ibh,Guo:2019joy}.   

Since leptons do not participate in the strong force and thus suffer negligible final-state interactions after being produced, they are widely considered as clean probes of the electromagnetic fields. The low mass dileptons ($M_{l\bar l}<1 ~ \rm GeV$) in heavy ion collisions come from the two body decays of the light vector mesons like $\rho,\omega,\phi$ and Dalitz decays of $\pi^0$ and $\eta$ mesons, commonly referred to as ``hadron cocktail''. The intermediate mass region of the dilepton spectra is governed by the open charm decay, and the high mass dileptons ($M_{l\bar l}>3 ~ \rm GeV$) are dominated by the initial-stage quark-antiquark annihilation, which is called Drell-Yan process, and heavy quarkonium decays. In high-energy nuclear collisions, a deconfined QCD matter is created, named quark-gluon plasma (QGP). The quark-antiquark annihilation in the QGP can also generate dileptons, which gives additional contribution to the dilepton spectra, especially in the low and intermediate mass regions~\cite{Braaten:1990wp,Rapp:2009yu,Rapp:2013ema,Linnyk:2015rco,Song:2018xca,Jorge:2025wwp}. The dilepton production in hot QCD medium under an external magnetic field is recently investigated, a considerable enhancement of dilepton production is found~\cite{Sadooghi:2016jyf,Bandyopadhyay:2017raf,Ghosh:2018xhh,Wang:2022jxx,Das:2021fma}. 

Although the strength of the electromagnetic fields is comparable to the QCD scale, the lifetime is very short~\cite{Deng:2012pc,Tuchin:2013ie,Wang:2021oqq,Yan:2021zjc,Chen:2021nxs}. This encourages the search for the electromagnetic fingerprint on hard probes, which are produced at the early stage of heavy ion collisions. Heavy flavor is one of such objectives. From the previous studies on the magnetic effects, the static properties, e.g. mass and shape, of open/hidden heavy flavor states are remarkably changed~\cite{Marasinghe:2011bt,Alford:2013jva,Cho:2014exa,Iwasaki:2021nrz,Guo:2015nsa,Chen:2020xsr}, the dynamic dissociation of quarkonium states in hot medium is significantly affected~\cite{Singh:2017nfa,Hasan:2018kvx,Hu:2022ofv}, and the heavy quark production in the initial stage is strongly enhanced at low transverse momentum~\cite{Chen:2024lmp,Chen:2024xjj}. In this paper, we focus on the magnetic field effect on the Drell-Yan process happened in the initial stage of heavy ion collisions. 

The paper is organized as follows. The framework is shown in Section~\ref{sec2}, in which we briefly review the Dirac equation in a magnetic field, and calculate the scattering amplitude and cross section of the Drell-Yan process. The calculation of the dilepton spectra and the comparison with the result in vacuum are shown in Section~\ref{sec3}. The conclusion is given in Section~\ref{sec4}.

\section{Theoretical framework}
\label{sec2}
\begin{figure}[!htb]
\includegraphics[width=0.35\textwidth]{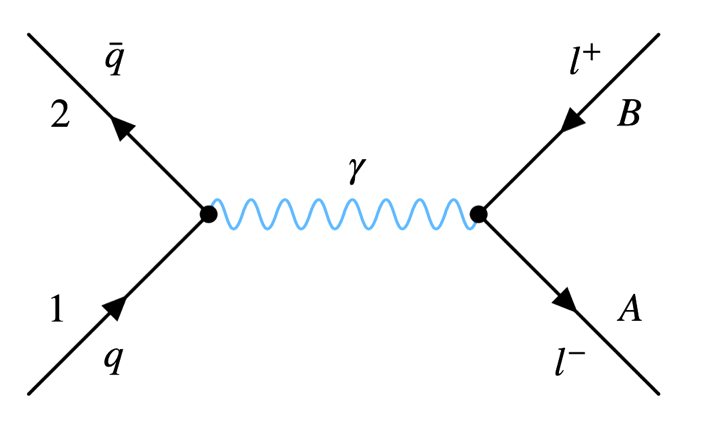}
\caption{The Feynman diagram of the Drell-Yan process $q\bar q\to l^+l^-$.}
\label{fig1}
\end{figure}

The Feynman diagram of the Drell-Yan process $q\bar q\to l^+l^-$ at leading order is shown in Fig.~\ref{fig1}, with a pair of incoming quarks, an exchanged photon and a pair of outgoing dilepton $e^+e^-$ or $\mu^+\mu^-$. 

The cross section of the process can be obtained by a perturbative pQCD calculation. The scattering amplitude in momentum space can be written as
\begin{eqnarray}
i{\mathcal M}=\delta_{ab}[\bar u(p_1)Q_qe\gamma_\mu v(p_2)]{-i\over k^2}[\bar v(p_B)(-e)\gamma_\mu u(p_A)],
\end{eqnarray}
where $\delta_{ab}$ is the color-conserving factor for the quark vertex, $-e$ and $Q_qe$ are the electric charge for electron and quarks with $Q_u=Q_c=2/3$ and $Q_d=Q_s=-1/3$. To compute the unpolarized cross section, we have to average (sum) over the spin and color degrees of freedom for the initial (final) particles. For massive quarks and leptons, the cross section in vacuum reads~\cite{Eichstaedt:2011ke,Ledwig:2017wit} 
\begin{eqnarray}
\sigma_{q\bar q\to l^+l^-}={4\pi \alpha_{\rm EM}^2 \over 3 s}{Q_q^2\over N_c}\sqrt{s-4m_q^2 \over s-4m_l^2}\left(1+{2m_q^2\over s} \right)\left(1+{2m_l^2\over s} \right),
\label{eq.woBwm}
\end{eqnarray}
where $N_c=3$ is the number of colors for quarks, $\alpha_{\rm EM}=1/137$ is the electromagnetic coupling constant, and $m_q$ and $m_l$ are the quark and lepton mass. If we neglect all the particle masses, the cross section becomes 
\begin{eqnarray}
\sigma_{q\bar q\to l^+l^-}={4\pi \alpha_{\rm EM}^2 \over 3 s}{Q_q^2\over N_c}.
\label{eq.woBwom}
\end{eqnarray}

We now consider the magnetic field effect. Since gluons and ghosts do not carry electric charge, their Feynman rules are not changed in an external magnetic field. The Feynman rules associated with quarks are controlled by the Dirac equation
\begin{equation}
\label{Dirac}
\left[i\gamma^\mu \left(\partial_\mu+iQ_qeA_\mu\right)-m\right]\psi=0,
\end{equation}
where $A_\mu$ is the electromagnetic potential. We consider an external magnetic field $B$ in the direction of $z$-axis and choose the Landau gauge with ${\bm A}=Bx{\bm e}_y$. In this case, the momentum along the $x$-axis is not conserved. Taking into account the Landau energy levels for a fermion moving in an external magnetic field, the quark energy square can be separated into a continuous part and a discrete part,
\begin{equation}
\label{Landau}
\epsilon^2 = p_z^2+\epsilon_n^2,\ \ \epsilon_n^2 = m^2+p_n^2,\ \ p_n^2=2n|Q_q|B,
\end{equation}
where $\epsilon_n$ and $p_n$ are the transverse energy and transverse momentum controlled by the quantum number $n = 0,1,2,\cdots$. The solution for the Dirac spinor can be written as~\cite{Melrose:1983svt,Kostenko:2018cgv,Kostenko:2019was},
\begin{equation}
\label{spinor}
\psi_{n,\sigma}^-(t,{\bm x}) = e^{-ip^\mu x_\mu} u_{n,\sigma}(x),\ \ \ \ \psi_{n,\sigma}^+(t,{\bm x}) = e^{ip^\mu x_\mu} v_{n,\sigma}(x)
\end{equation}
with positive-energy spinors
\begin{eqnarray}
\label{positive}
u_{n,-}(x) = \frac{1}{f_n}\left[\begin{matrix}-ip_z p_n \phi_{n-1}\\(\epsilon+\epsilon_n)(\epsilon_n+m)\phi_n\\-ip_n(\epsilon+\epsilon_n)\phi_{n-1}\\-p_z(\epsilon_n+m)\phi_n\end{matrix}\right],\ \ \ 
u_{n,+}(x) = \frac{1}{f_n}\left[\begin{matrix}(\epsilon+\epsilon_n)(\epsilon_n+m)\phi_{n-1}\\-ip_z p_n \phi_n\\p_z(\epsilon_n+m)\phi_{n-1}\\ip_n(\epsilon+\epsilon_n)\phi_n\end{matrix}\right]
\end{eqnarray}
and negative-energy spinors
\begin{eqnarray}
\label{negative}
v_{n,+}(x) = \frac{1}{f_n}\left[\begin{matrix}-p_n(\epsilon+\epsilon_n)\phi_{n-1}\\-ip_z(\epsilon_n+m)\phi_n\\-p_zp_n\phi_{n-1}\\i(\epsilon+\epsilon_n)(\epsilon_n+m)\phi_n\end{matrix}\right],\ \ \ 
v_{n,-}(x) = \frac{1}{f_n}\left[\begin{matrix}-ip_z(\epsilon_n+m)\phi_{n-1}\\-p_n(\epsilon+\epsilon_n)\phi_n \\-i(\epsilon+\epsilon_n)(\epsilon_n+m)\phi_{n-1}\\p_zp_n\phi_n\end{matrix}\right],
\end{eqnarray}
where $\sigma=\pm 1$ is the eigenvalue of the Pauli matrix, the four-momentum is defined as $p_\mu = (\epsilon,0,p_y,p_z)$ with $p_y=aQ_qB$ controlled by the center of gyration $a$, $f_n = 2\sqrt{\epsilon\epsilon_n(\epsilon_n+m)(\epsilon_n+\epsilon)}$ is the normalization constant, and $\phi_n$ is the harmonic-oscillator wave function,
\begin{equation}
\label{harmonic}
\phi_n(x-a) =\frac{1}{L\sqrt{\sqrt{\pi}\lambda 2^n n!}}H_n\left(\frac{x-a}{\lambda}\right)e^{-(x-a)^2/(2\lambda^2)}
\end{equation}
with the wave function width $\lambda=1/\sqrt{|Q_q|eB}$, the Hermite polynomial $H_n(x)$, and the normalization length $L$. Note that, while the spin eigenvalue $\sigma=\pm$ is used to describe the Dirac spinors, the spinors in normal case are not the spin eigenstates. Only for massless quarks, $u_{n,\sigma}$ and $v_{n,\sigma}$ become the spin eigenstates.   

With the known Dirac spinors, the scattering amplitude of the Drell-Yan process in an external magnetic field can be expressed as
\begin{eqnarray}
i\mathcal M &=& -iQ_qe^2\int d^4 x \int d^4 x' e^{ip_1^\rho x_\rho}\bar u_{n_1,\sigma_1}(x) \gamma_\mu e^{ip_2^\rho x_\rho} v_{n_2,\sigma_2}(x) \int \frac{d^4 k}{(2\pi)^4} \frac{-i g^{\mu\nu}e^{-ik^\rho(x-x')_\rho}}{k^2} e^{-i p_B^\rho x_\rho'}  \nonumber\\
&&\times \bar v_{n_B,\sigma_B}(x')\gamma_\nu e^{- i p_A^\mu x'_\mu} u_{n_A,\sigma_A}(x'),
\end{eqnarray}
where we have labeled the photon momentum with $k=(\omega,{\bm k})$. 

Since the Dirac spinors are only the functions of $x$, one can first integrate over $y,y',z,z',t,t'$, which gives the delta functions,
\begin{eqnarray}
\int d t \int d t'  e^{-i\epsilon_A t'}e^{-i\epsilon_Bt'}  e^{-i\omega(t-t')}e^{i \epsilon_1t} e^{i\epsilon_2 t} &=&(2\pi)^2\delta(\epsilon_1+\epsilon_2-\omega)\delta(\epsilon_A+\epsilon_B-\omega),\nonumber\\
\int d y \int d y'  e^{ip_{Ay} y'}e^{ip_{By}y'}  e^{ik_{y}(y-y')}e^{-i p_{1y}y} e^{-ip_{2y} y} &=&(2\pi)^2\delta(p_{1y}+p_{2y}-k_{y})\delta(p_{Ay}+p_{By}-k_{y}),\nonumber\\
\int d z \int d z'  e^{ip_{Az} z'}e^{ip_{Bz}z'}  e^{ik_{z}(z-z')}e^{-i p_{1z}z} e^{-ip_{2z} z} &=&(2\pi)^2\delta(p_{1z}+p_{2z}-k_{z})\delta(p_{Az}+p_{Bz}-k_{z}),
\end{eqnarray}
and then integrate over $k_y$, $k_z$, and $\omega$, the scattering amplitude becomes 
\begin{eqnarray}
\label{amplitude}
i\mathcal M &=& -Q_qe^2(2\pi)^2\int d x \int d x' \bar u_{n_1,\sigma_1}(x) \gamma_\mu  v_{n_2,\sigma_2}(x) \int d k_{x} \frac{g^{\mu\nu}e^{ik_{x}(x-x')}}{(\epsilon_A+\epsilon_B)^2-(p_{Az}+p_{Bz})^2-\frac{(a_A-a_B)^2}{{\lambda'}^4_B}-k_{x}^2} \nonumber\\
&&\times \bar v_{n_B,\sigma_B}(x')\gamma_\nu  u_{n_A,\sigma_A}(x')\delta(\epsilon_1+\epsilon_2-\epsilon_A-\epsilon_B)\delta(p_{1z}+p_{2z}-p_{Az}-p_{Bz})\delta \Big(\frac{a_1-a_2}{\lambda^2}-\frac{a_A-a_B}{{\lambda'}^2}\Big)
\end{eqnarray}
with $\lambda'=1/\sqrt{eB}$ for the final state leptons.

We first consider the contribution from the lowest Landau level (LLL), which is a good approximation for a very strong magnetic field and often used in literature~\cite{Singh:2017nfa,Hattori:2015aki,Chen:2024lmp}. Taking the quantum numbers $n_1=n_2=n_A=n_B=0$, the initial and final state Dirac spinors are reduced to 
\begin{eqnarray}
&& u_{0,-}(x) = \frac{1}{f_0}\left[\begin{matrix}0 \\ 2m_i(\epsilon_i+m_i)\phi_0(x-a_i)\\0\\-2m_ip_{i,z}\phi_0(x-a_i)\end{matrix}\right],\ \ \  
v_{0,+}(x) = \frac{1}{f_0}\left[\begin{matrix}0\\-i2m_ip_{i,z}\phi_0(x-a_i)\\0\\i2m_i(\epsilon_i+m_i)\phi_0(x-a_i)\end{matrix}\right],\nonumber\\
&& u_{0,+}(x)=v_{0,-}(x)=0.
\label{eq,spinors.lll}
\end{eqnarray}
In this case, the integration over $x$ and $x'$ in the scattering amplitude \eqref{amplitude} can be done analytically, 
\begin{eqnarray}
{\mathcal I}_1 &=&\int dx \bar u_{0,-}(x) \gamma_\mu  v_{0,+}(x) e^{ik_{x}x}\nonumber\\
&=&\frac{-i}{ 2L^2\sqrt{(\epsilon_1+m_q)\epsilon_1(\epsilon_2+m_q)\epsilon_2}}e^{-\frac{(a_1-a_2)^2}{4\lambda^2}+\frac{i(a_1+a_2)k_{x}}{2}-\frac{k_{x}^2\lambda^2}{4}}\nonumber\\
&&\times\left[p_{2z}(\epsilon_1+m_q)+p_{1z}(\epsilon_2+m_q),0,0,p_{1z}p_{2z}+(\epsilon_1+m_q)(\epsilon_2+m_q)\right]
\end{eqnarray}
and
\begin{eqnarray}
{\mathcal I}_2 &=& \int dx' \bar v_{0,+}(x')\gamma_\nu  u_{0,-}(x') e^{-ik_{x}x'}\nonumber\\
&=&\frac{i}{ 2L^2\sqrt{(\epsilon_A+m_l)\epsilon_A(\epsilon_B+m_l)\epsilon_B}}e^{-\frac{(a_A-a_B)^2}{4{\lambda'}^2}-\frac{i(a_A+a_B)k_{x}}{2}-\frac{k_{x}^2{\lambda'}^2}{4}}\nonumber\\
&&\times \left[p_{Bz}(\epsilon_A+m_l)+p_{Az}(\epsilon_B+m_l),0,0,p_{Az}p_{Bz}+(\epsilon_A+m_l)(\epsilon_B+m_l)\right],
\end{eqnarray}
and the amplitude can be simplified as
\begin{eqnarray}
i{\mathcal M} &=& -Q_qe^2\frac{1}{ 4L^4 \sqrt{(\epsilon_1+m_q)\epsilon_1(\epsilon_2+m_q)\epsilon_2(\epsilon_A+m_l)\epsilon_A(\epsilon_B+m_l)\epsilon_B}}\nonumber\\
&&\times\Big[\left(p_{2z}(\epsilon_1+m_q)+p_{1z}(\epsilon_2+m_q)\right)\left(p_{Bz}(\epsilon_A+m_l)+p_{Az}(\epsilon_B+m_l)\right)\nonumber\\
&&-\left(p_{1z}p_{2z}+(\epsilon_1+m_q)(\epsilon_2+m_q)\right)\left(p_{Az}p_{Bz}+(\epsilon_A+m_l)(\epsilon_B+m_l)\right)\Big]\nonumber\\
&&\times e^{-\frac{(a_1-a_2)^2}{4\lambda^2}}e^{-\frac{(a_A-a_B)^2}{4{\lambda'}^2}} {\mathcal I}(a_A,a_B,\epsilon_A,\epsilon_B,p_{Az},p_{Bz})\nonumber\\
&&\times (2\pi)^2\delta(\epsilon_1+\epsilon_2-\epsilon_A-\epsilon_B)\delta(p_{1z}+p_{2z}-p_{Az}-p_{Bz})\delta\left(\frac{a_1-a_2}{\lambda^2}-\frac{a_A-a_B}{{\lambda'}^2}\right),
\end{eqnarray}
where ${\mathcal I}$ is the principle value integral over $k_x$,
\begin{eqnarray}
{\mathcal I}(a_A,a_B,\epsilon_A,\epsilon_B,p_{Az},p_{Bz}) = \int dk_{x}\frac{e^{-\frac{i(a_A+a_B)k_{x}}{2}-\frac{k_{x}^2{\lambda'}^2}{4}+\frac{i(a_1+a_2)k_{x}}{2}-\frac{k_{x}^2\lambda^2}{4}}}{(\epsilon_A+\epsilon_B)^2-(p_{Az}+p_{Bz})^2-\frac{(a_A-a_B)^2}{{\lambda'}^4}-k_{x}^2}. 
\end{eqnarray}
When move to the center-of-mass frame with $p_{Az}=-p_{Bz}$, $\epsilon_A=\epsilon_B$, the principle value integral becomes
\begin{eqnarray}
{\mathcal I}(a_A,a_B,\epsilon_A) = \int dk_{x}\frac{e^{\frac{i(a_1+a_2-a_A-a_B)k_{x}}{2}-\frac{{\lambda'}^2+\lambda^2}{4}k_{x}^2}}{4\epsilon^2_A-\frac{(a_A-a_B)^2}{{\lambda'}^4}-k_{x}^2}. 
\label{eq.intkx}
\end{eqnarray}

An external magnetic field opens up a decay channel for converting a photon to a dilepton pair $\gamma\to {l'}^+{l'}^-$. The single-photon channel is consistent with the conservation of both momentum and energy, because the Lorentz invariance is broken in an external field. The conversion of a photon into a pair of ${l'}^+{l'}^-$ confined to the lowest Landau level has been investigated for many years~\cite{Wunner:1979ir,1980ApJ...238..296D}. The $\gamma\to {l'}^+{l'}^-$ cross section is strongly enhanced and larger than the two-photon annihilation process, when the magnetic field is strong enough. These results can probably be observed in magnetars and relativistic heavy ion collisions. 

The $\gamma\to {l'}^+{l'}^-$ process gives a finite width to the intermediate-state photon in the Drell-Yan process, when the photon energy exceeds the threshold $2m_{l'}$. The decay rate of a photon into a dilepton pair ${l'}^+{l'}^-$ at the lowest Landau level is given by~\cite{Kostenko:2018cgv}
\begin{eqnarray}
\Gamma (\omega,\theta) = 2 \alpha_{\rm EM}\frac{B}{B_{l'}}\frac{m_{l'}^4}{{\omega}^2\sin^2\theta(\omega^2\sin^2\theta-4m_{l'}^2)^{1/2}}e^{-B_{l'}/(2B)(\omega\sin\theta/m_{l'})^2}\sin\theta
\label{eq.decayratelll}
\end{eqnarray}
with $B_{l'} = m_{l'}^2/|e|$, where $\theta$ is the angle between the photon motion and the magnetic field. The intermediate photon can also decay into a quark-antiquark pair $\gamma \to q'\bar {q'}$, which will further reduce the Drell-Yan process. The decay width modifies the internal photon energy via $\omega \to \omega -i\Gamma/2$. In the center-of-mass frame with $\omega=2\epsilon_A$, the inverse of the photon propagator, namely the denominator of the principkle value integral \eqref{eq.intkx} becomes
\begin{eqnarray}
(2\epsilon_A-i{\Gamma}/{2})^2-\frac{(a_A-a_B)^2}{{\lambda'}^4}-k_{x}^2
\end{eqnarray}
which determines the pole of the photon propagator. To the first order in $\Gamma$, the pole is located at  
\begin{equation}
k_x = k_R+ik_I 
\end{equation}
with the real and imaginary parts,
\begin{eqnarray}
k_R^2&=&\frac{1}{2}\Big(4\epsilon_A^2-\frac{(a_A-a_B)^2}{{\lambda'}^4}\Big)+\frac{1}{2}\sqrt{\Big(4\epsilon_A^2-\frac{(a_A-a_B)^2}{{\lambda'}^4}\Big)^2+\epsilon_A^2 \Gamma^2},\nonumber\\
k_I^2&=&-\frac{1}{2}\Big(4\epsilon_A^2-\frac{(a_A-a_B)^2}{{\lambda'}^4}\Big)+\frac{1}{2}\sqrt{\Big(4\epsilon_A^2-\frac{(a_A-a_B)^2}{{\lambda'}^4}\Big)^2+\epsilon_A^2 \Gamma^2}.
\end{eqnarray}

Considering that the pole depends not on the gyration parameters $a_A$ and $a_B$ separately but on their relative distance $a_A-a_B$,  one can simply take $a_A=0$ and do only the integration over $a_B$ in the calculation of the cross section. In addition, in the strong magnetic field limit, the wave function widths $\lambda,\lambda'\to 0$, the exponent term $\exp[-({\lambda'}^2+\lambda^2)/4k_x^2]$ in the principle value integral \eqref{eq.intkx} can be dismissed. Taking the Residue theorem, the integral ${\mathcal I}$ and scattering amplitude ${\mathcal M}$ can be approximately expressed as  
\begin{eqnarray}
{\mathcal I} &=& i\pi sqn(a_1){e^{i|a_1|(k_R+ik_I)}\over k_R+ik_I},\nonumber\\
i{\mathcal M} &=& iQ_qe^2 {m_qm_l\over 2\pi L^4 \epsilon_A^2}{\mathcal I}e^{-\left({(a_1-a_2)^2\over 4\lambda^2}-{a_B^2\over 4{\lambda'}^2}\right)}(2\pi)^3\delta^3(p_1+p_2-p_A-p_B).
\end{eqnarray}
With the amplitude, we can calculate the cross section,
\begin{eqnarray}
\sigma = \frac{1}{|\beta_1-\beta_2|}\frac{L^3}{T}\int \frac{da_B}{L} \frac{da_1L}{2\pi{\lambda}^2} \frac{da_2L}{2\pi{\lambda}^2} \frac{dp_{1z} L}{2\pi} \frac{dp_{2z} L}{2\pi}|{\mathcal M}|^2
\end{eqnarray}
with initial particle velocities in the $z$-direction $\beta_1=p_{1z}/\epsilon_1$ and $\beta_2=p_{2z}/\epsilon_2$. Taking the relative velocity $|\beta_1-\beta_2|=2|p_{Az}|/\epsilon_A$ in the center-of-mass frame, the cross section becomes
\begin{eqnarray}
\sigma_{q\bar q\to l^+l^-} &=& \frac{(Q_qe^2)^2m_q^2m_l^2}{16\pi^3|p_{Az}|\epsilon_A^3\lambda^4}\int da_B da_1 da_2 dp_{1z}dp_{2z}e^{-\left({(a_1-a_2)^2\over 2\lambda^2}-{a_B^2\over 2{\lambda'}^2}\right)}|{\mathcal I}|^2\delta^3(p_1+p_2-p_A-p_B)\nonumber\\
&=& \frac{|Q_q|e^4 m_q^2m_l^2}{16\pi|p_{Az}||p_{1z}|\epsilon_A^3\Gamma}\left[\frac{\pi}{2}+\arctan\left(\frac{2\epsilon_A}{\Gamma}\right)\right].
\end{eqnarray}
Considering $p_{Az}=\sqrt{\epsilon_A^2-m_l^2},\ p_{1z}=\sqrt{\epsilon_1^2-m_q^2}$ and $\sqrt{s}=2\epsilon_A=2\epsilon_1$, the colliding energy dependence of the cross section can finally be written as 
\begin{eqnarray}
\sigma_{q\bar q\to l^+l^-}(\sqrt s)={32\pi \alpha_{\rm EM}^2 |Q_q| m_q^2m_l^2\over \sqrt{(s-4m_l^2)(s-4m_q^2)}\ s^2}{\sqrt{s}\over\Gamma}\left[{\pi \over 2}+\arctan\left(\frac{\sqrt{s}}{\Gamma}\right)\right],
\label{eq.csweB}
\end{eqnarray}
where the photon decay rate includes all the possible lepton channels $e^+e^-, \mu^+\mu^-, \tau^+\tau^-$ and quark channels $u\bar u, d\bar d, s\bar s, c\bar c$.
\begin{eqnarray}
\label{decay}
\Gamma (s,\theta) &=& \sum_i 2 \alpha_{\rm EM}|Q_i||eB|\frac{m_i^2\sin\theta}{s\sin^2\theta\sqrt{s\sin^2\theta-4m_i^2}}e^{-s\sin^2\theta/(2|Q_i|eB)},\nonumber\\
i &=& e,\mu,\tau,u,d,s,c.
\end{eqnarray}
We can see that, the magnetic field dependence of the cross section is hidden only in the decay width.

Since the magnetic field is defined in the $z$-direction, the Lorentz force acting on the initial and final state charged particles is in the transverse plane $x-y$. The momenta in the $z$-direction of the two initial (final) particles are $\pm p_z$, and the photon motion in the center-of-mass frame should be perpendicular to the magnetic field, which means $\theta=\pi/2$ in the decay rate. Therefore, the decay rate will diverge at the threshold $\sqrt{s}=2m_i$. The energy dependence of the cross section under a magnetic field, Eq.~\eqref{eq.csweB}, is very different from that in vacuum, Eq.~\eqref{eq.woBwom}. An important reason is the change in the size of the phase space. The LLL approximation in a strong magnetic field reduces the dimension of the phase space from three to two. This dimension reduction dramatically changes the momentum integration element. Due to the cancellation of the energy threshold factors from the decay rate $\Gamma$, there is no divergence for the cross section. It is nonzero at the scattering thresholds $\sqrt s=2m_q$ and $\sqrt s=2m_l$ and approaches zero at other decay thresholds $\sqrt s=2m_i\ (i\neq q,l)$.  

How good is the LLL approximation, and is it enough for the magnetic field created in nuclear collisions at LHC energy? To answer these questions we consider now the contribution from the next Landau level (NLL) and discuss the reliability of our calculation up to the NLL.  

In the calculation of the decay rate $\Gamma$ at NLL, at least one of the decayed fermions is at the NLL state, and for the cross section $\sigma$ at NLL, at least one of the fermions, including the incoming quarks, outgoing leptons and all possible decayed fermions, is at the NLL state. Including the contributions from the Landau levels $n=0$ and $n=1$, there exist different decay and scattering thresholds,
\begin{equation}
\sqrt{s} = 2m_i
\end{equation}
at LLL and 
\begin{equation} 
\sqrt{s} = m_i+\sqrt{m_i^2+2|Q_i|eB}, \ \ 2\sqrt{m_i^2+2|Q_i|eB}
\end{equation}
at NLL. The NLL contribution to the decay rate will introduce further divergence at these new thresholds. Similar to the LLL case, the cross section will be finite at all thresholds. 

At any Landau level $n=0,1,\cdots$, there are more than one decay threshold $s_i^{(n)} (i=e,\mu,\tau,u,d,s,c)$. Considering the particle electric charge and mass, the minimum threshold is controlled by the $d-$quark. Neglecting the $d-$quark mass $m_d$, the minimum threshold is
\begin{equation}
s_d^{(n)}=2n|Q_d|eB={2\over 3}n (eB).
\end{equation}  
In the energy region
\begin{equation}
s<s_d^{(n)},
\end{equation}
there are no contribution from the Landau levels $n,n+1,\cdots$. This means that, up to the Landau level $n-1$ the magnetic field effect is completely included in the calculation of the Drell-Yan process in the region $s<s_d^{(n)}$. The lowest Landau level approximation or the low Landau levels approximation is good, when the magnetic field is strong and/or the colliding energy is low. 
 
\section{Invariant mass spectra}
\label{sec3}
The differential cross section of the Drell-Yan process in a nucleon-nucleon collision can be written as
\begin{eqnarray}
{d\sigma_{q\bar q\to l^+l^-}^{pp}\over dM^2}&=&\int_0^1 dx_1d x_2 \sum_{q=u,d,s}\left[f_q(x_1,Q)f_{\bar q}(x_2,Q)+f_{\bar q}(x_1,Q)f_{q}(x_2,Q)\right]{d\hat \sigma_{q\bar q\to l^+l^-}\over dM^2},
\end{eqnarray}
where $\hat\sigma$ labels the above calculated cross section for the elemental process $q\bar q\to l^+ l^-$, and $f_i$ is the parton distribution function in a nucleon. In this study, we take the CTEQ6 parton distribution function~\cite{Nadolsky:2008zw}. The summation is over all possible initial quark-antiquark pairs, $x_1$ and $x_2$ are the longitudnial momentum fractions of quark and antiquark, $Q$ is the energy scale, and $M$ is the invariant mass of the lepton pair $l^+l^-$.
Writing the momenta of the initial quark and antiquark as
\begin{eqnarray}
p_1={\sqrt{s}\over 2}(x_1,0,0,x_1), ~ \ p_2={\sqrt{s}\over 2}(x_2,0,0,-x_2),
\end{eqnarray}
where $\sqrt{s}$ is the colliding energy of p-p, the quark-antiquark colliding energy $\hat s$ can be expressed as 
\begin{equation}
\hat s = x_1x_2s=M^2,\ \ x_1 = {M\over \sqrt{s}}e^y,\ \ x_2 = {M\over \sqrt{s}}e^{-y}
\end{equation}
with $y$ being the longitudinal rapidity for the pair. Since the Drell-Yan process is in $s-$channel, one can take $Q^2=M^2$. Using the definition of rapidity $y=(1/2)\ln(x_1/x_2)$, the double differential cross section in vacuum can be expressed as
\begin{eqnarray}
{d^2\sigma_{q\bar q\to l^+l^-}^{pp}\over dMdy}(s,M)={8\pi \alpha_{\rm EM}^2 \over 9 M}{1\over s}\sum_{q=u,d,s}Q_q^2\left[f_q(x_1)f_{\bar q}(x_2)+f_{\bar q}(x_1)f_{q}(x_2)\right]
\end{eqnarray}
with which the experimental observable can be calculated, 
\begin{equation}
{d\sigma_{q\bar q\to l^+l^-}^{pp}\over dM}=\int {d^2\sigma_{q\bar q\to l^+l^-}^{pp}\over dMdy}dy.
\end{equation}
The spectra under a magnetic field can be calculated in a similar way, the field effect is included in the elementary cross section $\hat\sigma$.
\begin{figure}[!htb]
\centering
\includegraphics[width=0.7\textwidth]{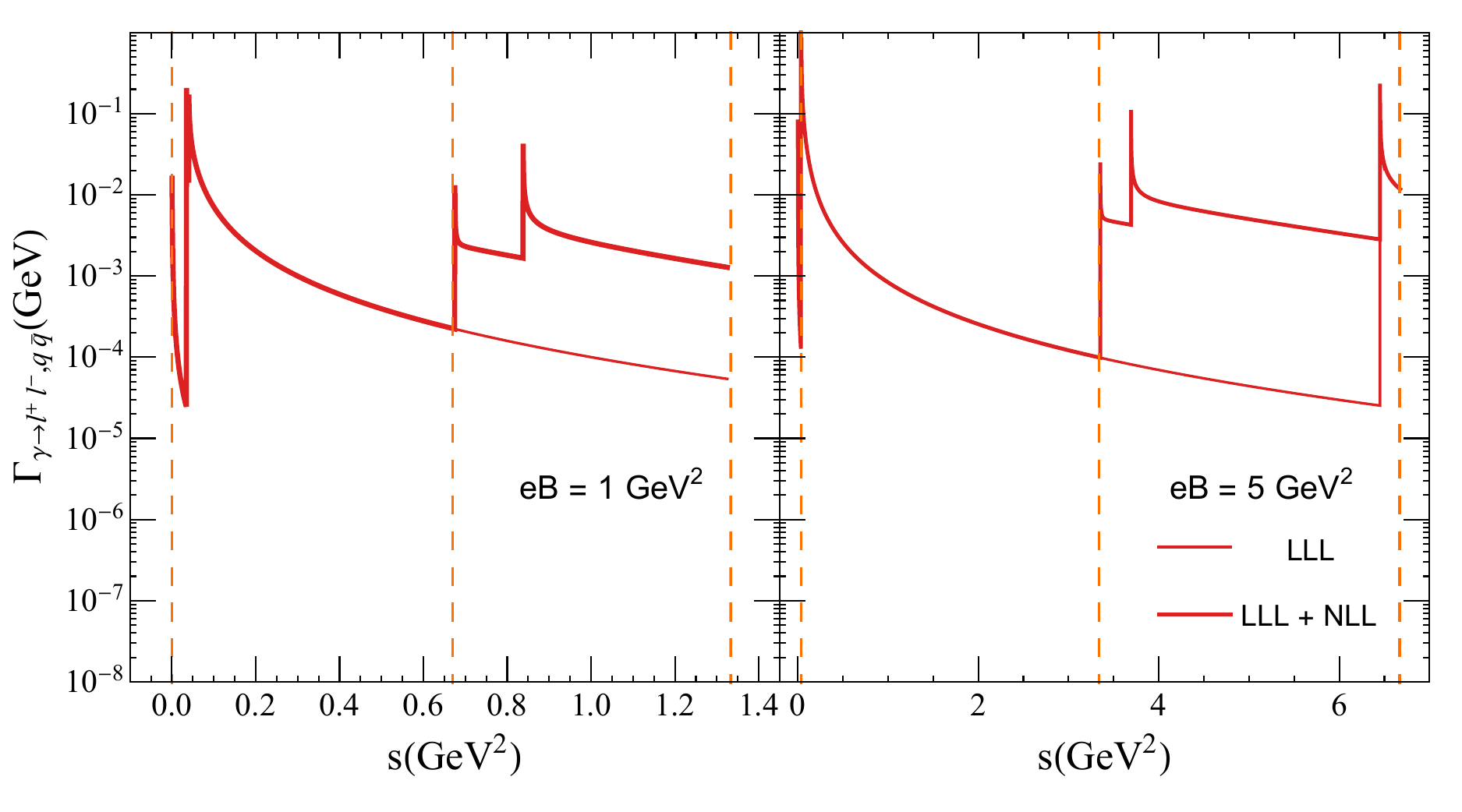}
\caption{The total photon decay rate including the decay channels $\gamma\to e^+e^-,\ \mu^+\mu^-,\ \tau^+\tau^-,\ u\bar u,\ d\bar d,\ s\bar s,\ c\bar c$ in an external magnetic field $eB=$ 1 (left panel) and 5 (right panel) GeV$^2$. The thin and thick solid lines represent the calculations up to LLL and NLL, and the three vertical dashed lines from left to right indicate the locations of the minimum thresholds $2/3n|eB|$ at Landau levels $n=0,\ 1,\ 2$. }
\label{fig2}
\end{figure}

A strong magnetic field can be created in non-central heavy ion collisions. To investigate the influence of the magnetic field on the Drell-Yan process, we study the dilepton spectra in heavy ion collisions. The production cross section in a nucleus-nucleus (AB) collision with impact parameter ${\bm b}$ can be written as a superposition of the proton-proton collisions,
\begin{eqnarray}
{d^2\sigma_{q\bar q\to l^+l^-}^{AB} \over dMdy}=\int d^2{\bm x}_tdz_A dz_B \rho_A( {\bm x}_t+{{\bm b}\over2},z_A)\rho_B( {\bm x}_t-{{\bm b}\over 2},z_B) \mathcal {R}^A_q \mathcal {R}^B_q{d^2\sigma_{q\bar q\to l^+l^-}^{pp}\over dMdy},
\end{eqnarray}
where $\rho_A$ and $\rho_B$ are the nucleon distributions in the nucleus $A$ and $B$, which can be taken as the Woods–Saxon function~\cite{Miller:2007ri}. ${\bm x}_t\pm{\bm b}/2$ and $z_A (z_B)$ are the colliding nucleon's transverse and longitudinal coordinates in $A$ and $B$. The nuclear shadowing effect for quarks is embedded in the factors $\mathcal R_q^{A,B}$~\cite{Helenius:2012wd}. To simplify the numerical calculation, we neglect here the shadowing effect and take $\mathcal R_q^A=\mathcal R_q^B=1$. Under this approximation, the differential cross section can be simplified as
 \begin{eqnarray}
{d^2N_{q\bar q\to l^+l^-}^{AB} \over dMdy} = {T_{AB}({\bm b})\over \sigma_{\rm inel}}{d^2\sigma_{q\bar q\to l^+l^-}^{pp}\over dMdy},
\label{eq.spectAA}
\end{eqnarray}
where $T_{AB}({\bm b})=\int T_{A}({\bm x}_t-{{\bm b}/2})T_B({\bm x}_t+{{\bm b}/2})d^2{\bm x}_t$ with the thickness functions $T_A$ and $T_B$ in nucleus $A$ and $B$ describes the nuclear geometry~\cite{Miller:2007ri}, and $\sigma_{\rm inel}=71\ \rm mb$ is the inelastic scattering cross section at $\sqrt{s_{\rm NN}}=5.02\ \rm TeV$~\cite{ALICE:2012fjm}.

In our numerical calculation, we take into account the Landau levels up to $n=1$. In this case, the magnetic field effect is completely considered in the energy region $s<s_d^{(2)}$. Considering the possibly realized magnetic field in physics systems, we take $|eB|=1,\ 5$ GeV$^2$, that means the maximum energy for the complete calculation
\begin{equation}
s_d^{(2)}=4/3 eB=\left\{\begin{array}{ll}
4/3\ \textrm{GeV}^2 & \textrm{for}\  eB=1\ \textrm{GeV}^2\\
20/3\ \textrm{GeV}^2 & \textrm{for}\ eB=5\ \textrm{GeV}^2\end{array}\right. 
\end{equation}
The lepton and quark masses are taken from the Particle Data Group~\cite{ParticleDataGroup:2024cfk}, $m_e=0.51$ MeV, $m_\mu=100$ MeV, $m_\tau=1.78$ GeV, $m_u=2$ MeV, $m_d=5$ MeV, $m_s=93.5$ MeV and $m_c=1.27$ GeV. 

The total decay rate $\Gamma_{\gamma\to l^+l^-,q\bar q}$ up to LLL (thin solid lines) and NLL (thick solid lines) is shown in Fig.~\ref{fig2} for magnetic field $eB=$ 1 (left panel) and 5 (right panel) GeV$^2$. The rate diverges at any threshold for the opening of a new dacay channel at Landau level $n=0$ or for the sudden expansion of the phase space at Landau levels $n>0$. The locations of the minimum thresholds $s_d^{(n)}$ for Landau levels $n=0,1,2$ are indicated by three vertical dashed lines from left to right. In the energy region $s<s_d^{(1)}=2/3 eB$ (the second dashed line) the magnetic field effect at LLL is already complete, the two lines with LLL and LLL+NLL coincide, and there is no contribution from the Landau levels $n>0$. The contribution from NLL starts at $s_d^{(1)}$, it enlarges the region for full calculation to $s<s_d^{(2)}=4/3 eB$ (the third dashed line). At higher energy, the contribution from higher Landau levels $n>1$ is needed. With the rapid expansion of the phase space by increasing the Landau level, the decay rate increases significantly. Note that, the thresholds $s_i^{(0)}$ for $i=e,u,d$ are very close to zero.   
\begin{figure}
    \centering
\includegraphics[width=0.7\linewidth]{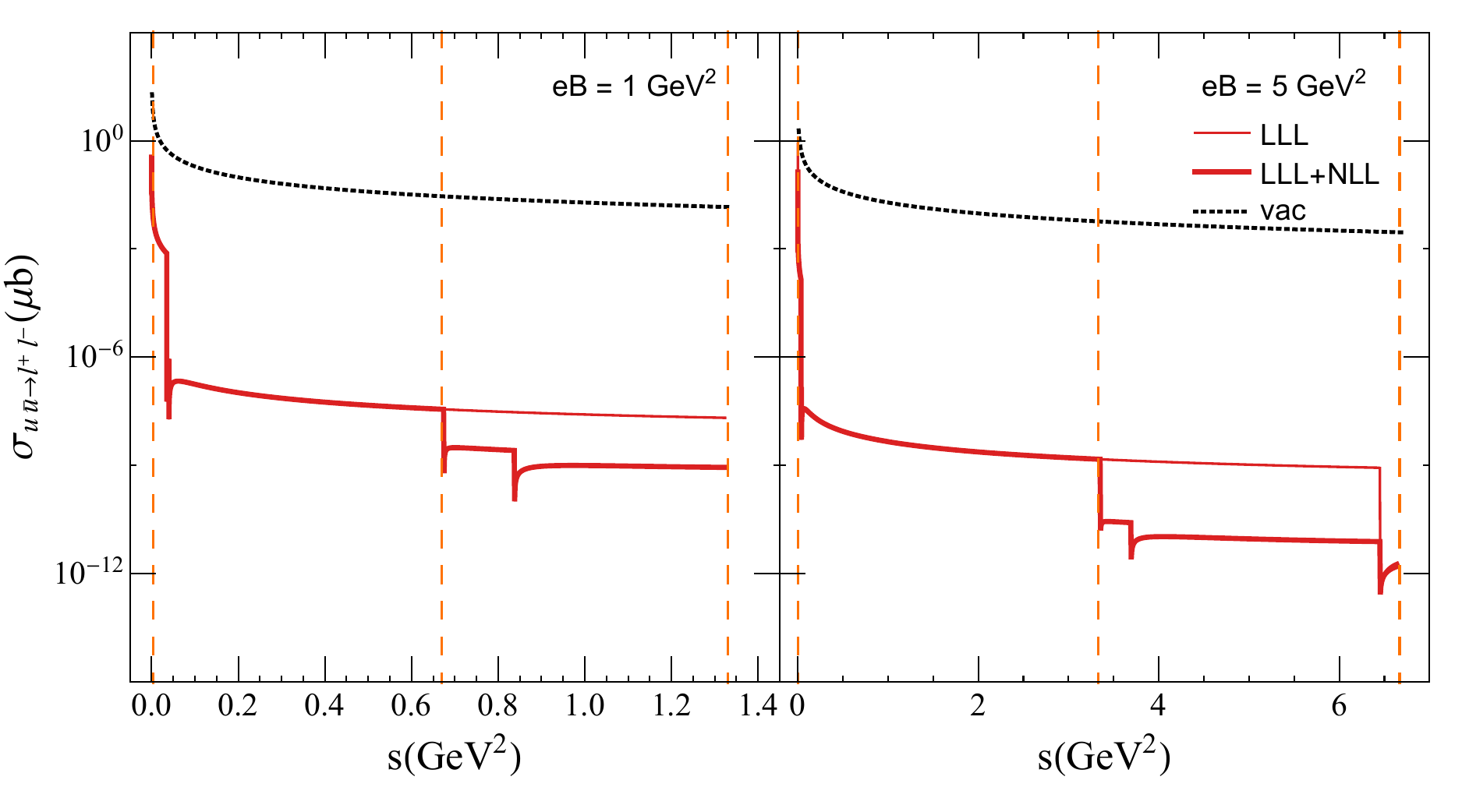}
    \caption{The Drell-Yan cross section $\sigma_{u\bar u\to l^+l^-}\ (l=e, \mu, \tau)$ in an external magnetic field $eB=$ 1 (left panel) and 5 (right panel) GeV$^2$. The thin and thick solid lines represent the calculations up to LLL and NLL, the dotted lines are the result in vacuum without magnetic field, and the three vertical dashed lines from left to right indicate the locations of the minimum thresholds $2/3n(eB)$ at Landau levels $n=0,\ 1,\ 2$.}
    \label{fig3}
\end{figure}

The cross section for the Drell-Yan processes $u\bar u\to l^+l^-\ (l=e, \mu, \tau)$ under an external magnetic field $eB=$1 (left panel),\ 5 (right panel) GeV$^2$ and the comparison with the result in vacuum (dotted lines) are shown in Fig.~\ref{fig3}. Again the calculations up to LLL (thin solid lines) and NLL (thick solid lines) are already complete in the energy regions $s<s_d^{(1)}$ and $s<s_d^{(2)}$. Compared to the result in vacuum without magnetic field, the cross section is dramatically suppressed by the photon decays $\gamma\to {l'}^+{l'}^-$ and $\gamma\to q\bar q$ in a strong magnetic field. For $q\neq u$ and $l'\neq l$, the cross section even drops down to zero at the corresponding thresholds. The suppression is mainly from the decay channels to light quark pairs and light lepton pairs, such decays lead to a strong suppression already at very low energy. Including the contribution from NLL, the increased decay rate shown in Fig.~\ref{fig2} leads to a further cross section suppression in the region $s_d^{(1)}<s<s_d^{(2)}$.       

\begin{figure}
	\centering
	\includegraphics[width=0.7\linewidth]{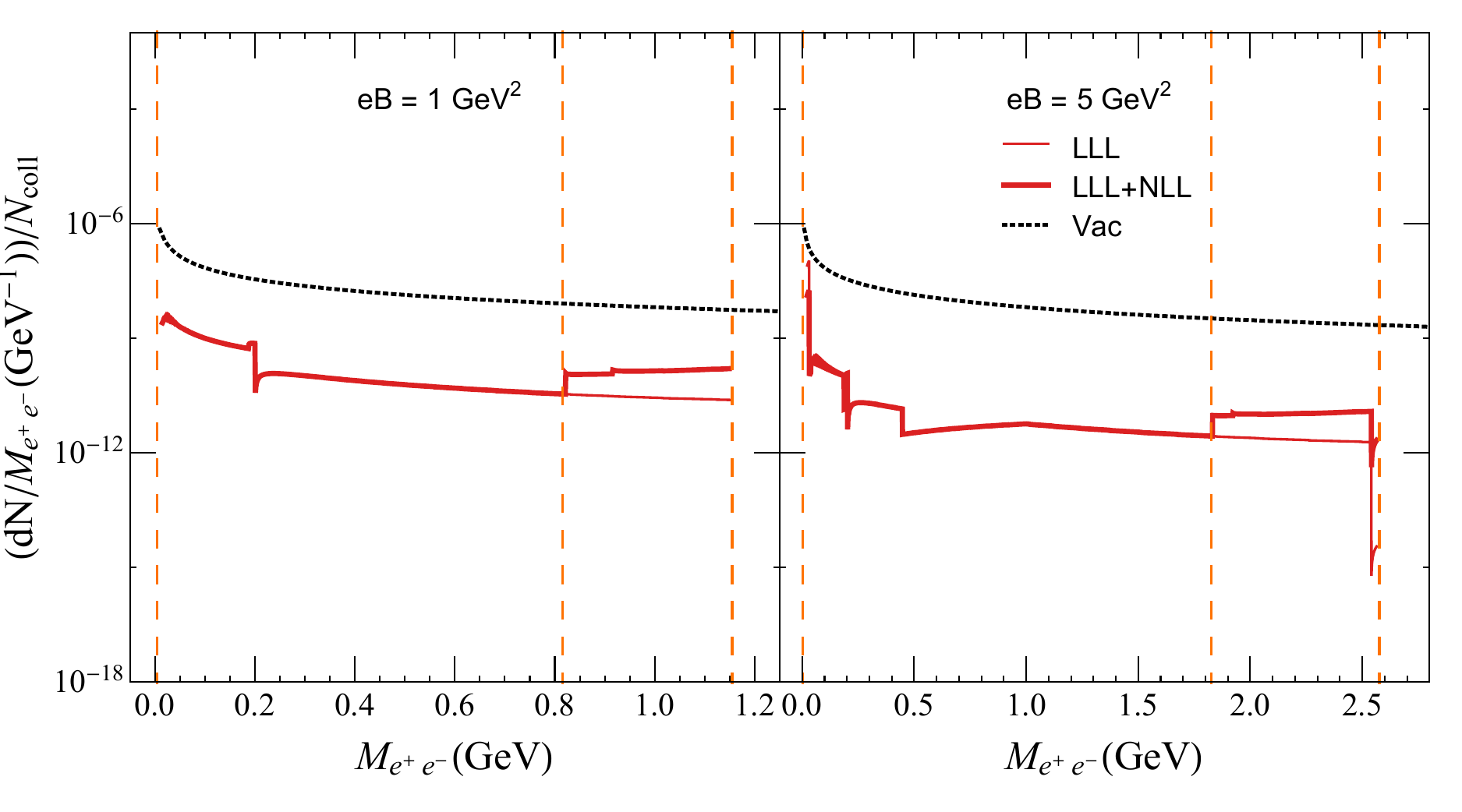}
	\caption{The dielectron spectrum $(dN/dM_{e^+e^-})/N_{coll}$ from Drell-Yan process $q\bar q\to e^+e^- (q=u,\ d,\ s)$ in a Pb-Pb collision at colliding energy $\sqrt{s_{\rm NN}}$ = 5.02 TeV in an external magnetic field $eB=$ 1 (left panel) and 5 (right panel) GeV$^2$. The thin and thick solid lines represent the calculations up to LLL and NLL, the dotted lines are the result in vacuum without magnetic field, and the three vertical dashed lines from left to right indicate the locations of the minimum thresholds $\sqrt{2/3n(eB)}$ at Landau levels $n=0,\ 1,\ 2$.}
	\label{fig4}
\end{figure}

We finally calculate the dielectron spectrum $(dN/dM_{e^+e^-})/N_{coll}$ for the Drell-Yan process $q\bar q\to e^+e^-\ (q=u,\ d,\ s)$ in a Pb-Pb collision at colliding energy $\sqrt{s_{\rm NN}}$ = 5.02 TeV with number of binary nucleon-nucleon collisions $N_{coll}$. The calculations up to LLL (thin solid lines) and NLL (thick solid lines) under an external magnetic field $eB$=1 (left panel) and 5 (right panel) GeV$^2$ and the comparison with the result in vacuum (dotted lines) are shown in Fig.~\ref{fig4}. In comparison with the cross section shown in Fig.~\ref{fig3}, the parton distribution $f_q$ which depends on $s$ largely changes the energy dependence of the spectrum. While the magnetic field suppresses the spectrum very strongly, the enlarged phase space at NLL makes the difference between with and without magnetic field smaller. This leads it possible to have a Drell-Yan enhancement at higher mass region where the higher Landau levels become important.    
\section{Summary}
\label{sec4}
We studied the Drell-Yan process in an external magnetic field. The field modifies not only the initial quark pair and final lepton pair through the solution of the Dirac equation for charged fermions but also the internal photon although it does not carry electric charge. The magnetic field opens the photon decay channels to dilepton and quark-antiquark pairs. Since an external field breaks down the translation invariance of the system, such decay process which is forbidden in vacuum satisfies both the energy and momentum conservation in the field.

We analyzed the applicability of the often used low Landau levels approximation. By considered the minimum energy threshold for the photon decay process at a fixed Landau level $n$, the magnetic field effect on the Drell-Yan process up to the Landau level $n-1$ is already complete in the energy region $s<2/3n(eB)$. The higher Landau levels do not contribute to this low energy region.  

We observed many steps in the photon rate, dilepton production cross section and dilepton mass spectrum, which are attributed to the emergence of the novel photon decay channels and the sudden expansion of the phase space. The decays happened in the magnetic field, especially to those light quark pairs and lepton pairs, significantly suppress the final state dilepton yield in the low and intermediate mass region, and the enlarged phase space at higher Landau levels may enhance the yield in high mass region. This can be tested in the future heavy ion experiment at LHC. 

\vspace{1cm}
\noindent {\bf Acknowledgement}: The authors acknowledge useful discussions with Shuzhe Shi, Taesoo Song, and Elena Bratkovskaya. S.Chen and P.Zhuang are supported by the Guangdong Major Project of Basic and Applied Basic Research No.2020B0301030008. J.Zhao is funded by the Deutsche Forschungsgemeinschaft (DFG, German Research Foundation) through the grant CRC-TR 211 ’Strong-interaction matter under extreme conditions’ - Project number 315477589 - TRR 211.

\bibliographystyle{apsrev4-1.bst}
\bibliography{Ref}

\end{document}